\documentclass[aps,pre,groupedaddress,amsmath,amssymb,10pt]{revtex4}
\usepackage[paperwidth=210mm,paperheight=297mm,centering,hmargin=3cm,vmargin=2.5cm]{geometry}
\usepackage{graphicx,xcolor}  
\usepackage{dcolumn}   
\usepackage{bm}        
\usepackage{verbatim}   
\usepackage{soul} 
\usepackage{enumitem}
\usepackage{tikz}

\newcommand{\rom}[1]{%
  \textup{\uppercase\expandafter{\romannumeral#1}}%
}

\begin{document}

\title{Periodic orbits exhibit oblique stripe patterns in plane Couette flow}
\author{Florian Reetz}
\affiliation{
Emergent Complexity in Physical Systems Laboratory (ECPS),\\
\'Ecole Polytechnique F\'ed\'erale de Lausanne, CH 1015 Lausanne, Switzerland
   }
   \author{Tobias M. Schneider}\thanks{email address for correspondence: tobias.schneider@epfl.ch}
\affiliation{
Emergent Complexity in Physical Systems Laboratory (ECPS),\\
\'Ecole Polytechnique F\'ed\'erale de Lausanne, CH 1015 Lausanne, Switzerland
   }
\date{\today}
\begin{abstract}

Spatio-temporally chaotic dynamics of transitional plane Couette flow may give rise to regular turbulent-laminar stripe patterns with a large-scale pattern wavelength and an oblique orientation relative to the laminar flow direction. A recent dynamical systems analysis of the oblique stripe pattern demonstrated that the Navier-Stokes equations have unstable equilibrium solutions that capture the three-dimensional spatial structure of the oblique stripe patterns. Here we identify unstable periodic orbits that not only show oblique large-scale amplitude modulation but also have a characteristic time-evolution, unlike steady equilibrium solutions. The periodic orbits represent standing waves that slowly propagate across wavy velocity streaks in the flow on viscous diffusion time scales. The unstable periodic orbits are embedded in the edge of chaos in a symmetry subspace of plane Couette flow and thereby may mediate transition to and from turbulent flows with oblique patterns.

\end{abstract}

\maketitle

\section{Introduction}

Transitional turbulence in linearly stable wall-bounded shear flows may segregate into laminar and turbulent regions. The spatial coexistence of turbulent and laminar flow has been observed in pipes with one extended space dimension \cite{Barkley2016} and in planar shear flows with two extended space dimensions \cite{Tuckerman2020}. The properties of turbulent-laminar coexistence have been studied extensively, both, in terms of statistical laws \cite{Lemoult2014,Chantry2017} and in terms of the dynamics at the interfaces between laminar and turbulent regions \cite{Barkley2011,Duguet2013}. For intermediate strength of driving, turbulent and laminar regions in wall-bounded shear flows may self-organize into a regular pattern of spatially periodic stripes or bands oriented obliquely relative to the laminar flow direction. Oblique stripe patterns have been observed experimentally and numerically in various wall-bounded shear flows suggesting universal mechanisms that create the regular pattern in the flow \cite{Manneville2017}. How turbulent-laminar stripes emerge at particular pattern wavelengths and particular oblique orientations remains an open problem.

The present article discusses regular oblique stripe patterns in plane Couette flow (PCF) where two extended parallel walls slide into opposite direction and drive an incompressible flow in the gap between the walls. First experimental observations of oblique stripe patterns in spatially extended PCF report a regular pattern at a Reynolds numbers of $\mathrm{Re}=U_w h/\nu=358$ \cite{Prigent2002,Prigent2003}. In PCF, $\mathrm{Re}$ is the only control parameter of the flow where $2U_w$ is the relative wall velocity, $2h$ is the gap-height and $\nu$ is the kinematic viscosity. The wavelength of the stripe pattern $\lambda$ is large compared to the gap height, $40h\lessapprox \lambda \lessapprox 60h$, and the observed orientation angle $\theta$ against the streamwise direction varies between $20^{\circ}\lessapprox \theta \lessapprox 40^{\circ}$. Oblique stripe patterns with similar ranges of $\lambda$ and $\theta$ have also been reproduced in direct numerical simulations of PCF at $330 \le \mathrm{Re} \le 380$ \cite{Duguet2010}. Both, experiments and simulations indicate a coupling of the pattern wavelength $\lambda$, the orientation angle $\theta$ and the Reynolds number $\mathrm{Re}$, such that $\lambda$ and $\theta$ tend to increase with decreasing $\mathrm{Re}$. A simple approximate relation, $\mathrm{Re}\sin(\theta)\approx\pi\lambda$, has been proposed based on a mean flow analysis of the stripe pattern \cite{Barkley2007}. Further empirical and theoretical studies are needed to more precisely describe the relation between wavelength $\lambda$, angle $\theta$ and the Reynolds number $\mathrm{Re}$.

Oblique turbulent-laminar stripes posses a mean flow that is invariant under continuous translations in the direction along the oblique stripes. Moreover, the mean flow is symmetric with respect to discrete translations with periodicity $\lambda$ across the pattern. These translation symmetries in two orthogonal space dimensions can be exploited in direct numerical simulations of flows showing the oblique stripes by using minimal periodic domains \cite{Barkley2005,Barkley2007}. In general, minimal periodic domains treat dimensions with translation symmetries as periodic. This minimizes the computational cost to study complex patterns and may simplify the simulated dynamics by disallowing instabilities breaking the imposed periodicity \citep{Golubitsky2002}. The minimal domain capturing the pattern contains a single period of the sustained periodic flow structure \cite{Jimenez1991}. To capture oblique turbulent-laminar stripe patterns, the two lateral dimensions of the minimal periodic domain must coincide with the translation-invariant directions of the pattern implying that the domain is tilted against the direction of the wall-velocity by angle $\theta$. In one direction the domain has an extension corresponding to the pattern wavelength $\lambda$ \cite{Barkley2005}. Thus in this minimal domain approach, not only $\mathrm{Re}$ but also $\theta$ and $\lambda$ are imposed parameters. This is in contrast to experiments and simulations in large extended domains where the angle $\theta$ and the wavelength $\lambda$ are unconstrained and can be freely selected by the flow. By choosing a tilted minimal periodic domain matching experimental observations, Barkley and Tuckerman \cite{Barkley2007} obtained well-converged temporal statistics of a stripe pattern at $\mathrm{Re}=350$, $\theta=24^{\circ}$ and $\lambda=40h$. They find the pattern's mean flow to be well-approximated by few harmonic functions with centro-symmetry about the center points of the laminar and the turbulent flow region, respectively. Oblique stripe patterns have also been numerically studied in tilted rectangular minimal domains of pressure-driven channel flow \cite{Tuckerman2014a} and in tilted non-rectangular minimal domains of PCF \cite{Deguchi2015a}.

When lowering $\mathrm{Re}$ from the turbulent regime towards the parameters where stripes are observed, the oblique pattern emerges from statistically homogeneous turbulence. Some studies have suggested that turbulent-laminar stripes are the consequence of a large-wavelength instability of unpatterned turbulence, in analogy to linear instabilities forming steady patterns described by non-linear amplitude equations \cite{Prigent2002,Manneville2012}. The identification of such large-wavelength instabilities is however challenging because the nonlinear dynamics of turbulent flows are chaotic in time and space \cite{Philip2011}. Consequently, there is no time-independent base state whose stability can be analyzed in a straightforward way. One way to disentangle the temporal dynamics from the spatial structure of the flow is to study exact equilibrium solutions of the governing equations with steady, time-invariant dynamics that capture the coexistence of different pattern motifs in space \cite{Knobloch2015}. If an exact equilibrium solution of the fully nonlinear Navier-Stokes equations captures the spatial coexistence of non-trivial `turbulent' flow structures and the laminar solution, then the turbulent-laminar coexistence may be described as a homoclinic connection in space between laminar flow and an equilibrium solution underlying a turbulent state. This description assumes that the spatial coordinate across a one-dimensional pattern is treated as a time-like variable \cite{Burke2006}. Such spatial homoclinic orbits have been identifed in PCF between laminar flow and spanwise localized wavy-streaky structures \cite{Schneider2010,Gibson2016,Salewski2019}. However, these time-invariant solutions do not capture the oblique orientation of the stripe pattern. Furthermore, these solutions do not suggest any particular pattern wavelength because they exist for arbitrary spanwise extent. Recently, an equilibrium solution of PCF has been found to capture the pattern of oblique turbulent-laminar stripes \cite{Reetz2019a}. This stripe equilibrium is computed at $\mathrm{Re}=350$ in the same tilted minimal periodic domain with $\theta=24^{\circ}$ and $\lambda=40h$ used to simulate oblique stripes \cite{Barkley2007}. The stripe equilibrium bifurcates first with a period of $\lambda=20h$ from the well-known Nagata equilibrium with wavy-streaky flow structures \cite{Nagata1990,Busse1992,Waleffe1997}, and second, increases the spatial period to $\lambda=40h$ in a spatial period-doubling bifurcation. This bifurcation sequence confirms the existence of a large-wavelength instability creating the oblique stripe patterns in PCF at a particular wavelength. 

The stripe equilibrium solution has the spatial structure of the oblique stripe pattern but is time-independent. Thus, the equilibrium cannot capture the temporal dynamics of oblique stripe patterns. The temporal dynamics of weakly turbulent shear flows has been described in terms of dynamically unstable invariant solutions including equilibrium solutions, but more importantly periodic orbits \cite{Kerswell2005,Eckhardt2007,Kawahara2012}. Unstable periodic orbits (UPO) have exactly periodic time evolution. These orbits capture recurrent motion of the flow and are thus particular interesting building blocks for the temporal dynamics \cite{Cvitanovic2013}. A sufficiently large number of UPOs may allow to quantitatively predict turbulent statistics \cite{Cvitanovic1991,Chandler2013}. Their bifurcations control changes in the statistics with system parameters \cite{Kreilos2014}. Also individual UPOs provide significant insights into the dynamics as they may capture characteristic time-evolution of the flow \cite{Kreilos2013a}. To capture the temporal dynamics of oblique stripe patterns in PCF, unstable periodic orbits are needed that both, have relevant intrinsic temporal dynamics and capture the spatial characteristics of the pattern. Such orbits have not been identified yet.

Unstable invariant solutions may transiently guide turbulent dynamics along their stable and unstable manifolds in state space \cite{Suri2017}, the space spanned by all solenoidal velocity fields \cite{chaosbook}. Specifically relevant for the dynamics of transition between laminar and turbulent flow are so-called \emph{edge states} \cite{Schneider2007b,DeLozar2012}. Edge states are attractive invariant sets within the \emph{edge of chaos}, a codimension-1 manifold in state space that separates trajectories approaching turbulent states from trajectories approaching the laminar solution. This property allows to identify edge states using a method known as \emph{edge-tracking} \cite{Skufca2006}. Many spatially localized invariant solutions in subcritical shear flows have been found because they are edge states for appropriately chosen flow parameters \cite{Schneider2010a,Avila2013,Khapko2013}. The stripe equilibrium is not an edge state \cite{Reetz2019a} and the edge of chaos for turbulent-laminar stripes has not been studied.

Following the recent identification of an equilibrium solution capturing the spatial structure of oblique turbulent-laminar stripes in PCF \cite{Reetz2019a}, in this article we discuss UPOs underlying oblique stripe patterns. Imposing discrete symmetries allows to identify one UPO in the edge of chaos at $\mathrm{Re}=350$ in a tilted domain. Continuation reveals two additional UPOs at the same $\mathrm{Re}$ connected via fold bifurcations. The UPOs represent standing wave oscillations with a wavelength of $\lambda=20$ and an oblique orientation at $\theta=24^{\circ}$. We describe the turbulent dynamics at $\mathrm{Re}=350$ and demonstrate the dynamical relevance of the identified UPOs both for the edge dynamics and the turbulent decay.

This article is structured as follows. To identify the symmetry subspace in which edge-tracking yields a UPO, we reduce the complexity of the spatio-temporal dynamics by imposing symmetries on the flow dynamics (Section \ref{sec:p4:dns}). Edge-tracking in two symmetry subspaces yields an attractor that is chaotic in one case but near-periodic in the other (Section \ref{sec:p4:edge}). For the near-periodic case, we identify a UPO that is connected to two other UPOs at $\mathrm{Re}=350$ via fold bifurcations (Section \ref{sec:p4:orbits}). The dynamical relevance of the three identified UPOs is discussed in Section \ref{sec:p4:discussion}.

\section{Numerical simulations in symmetry subspaces}\label{sec:p4:dns}
\begin{figure}[tb]
        \begin{tikzpicture}
    	\draw (0, 0) node[inner sep=0]{\includegraphics[width=0.89\linewidth,trim={0.0cm 5.3cm 0.0cm 5.0cm},clip]{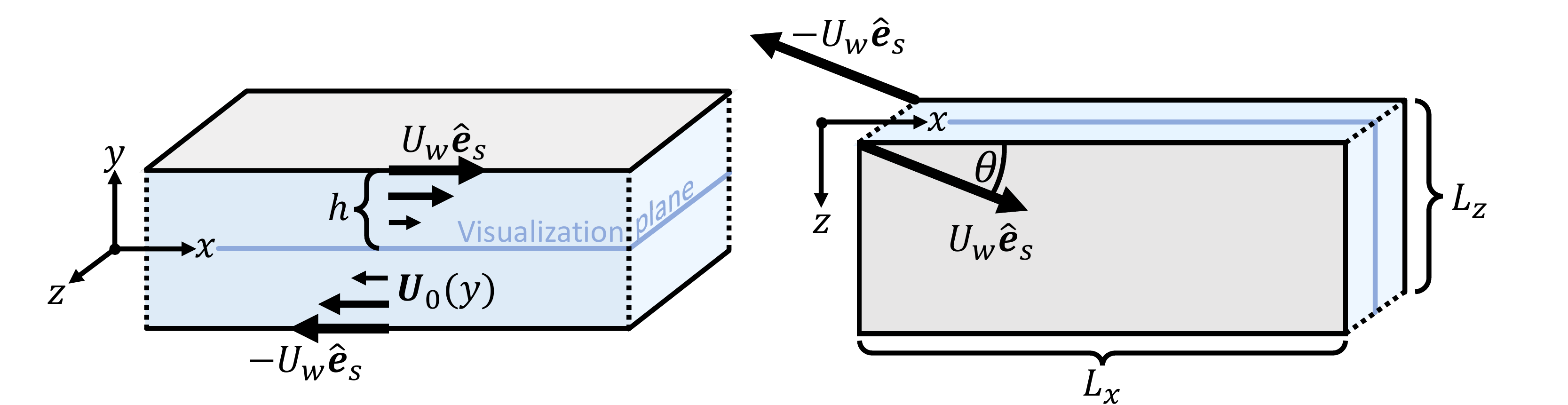}};
    	\draw (-6,-1.6) node {\textbf{(a)}};
    	\draw (0.2,-1.6) node {\textbf{(b)}};
		\end{tikzpicture}
\caption{\label{fig:pcf:systemGeometry} Geometry of plane Couette flow (PCF). A Newtonian fluid (light blue) is studied numerically in $x$-$z$-periodic domains, bounded in $y$ by two parallel walls (light grey). The walls move with a relative velocity of $2U_w$ in opposite directions and thereby drive a flow. The unit vector $\hat{\bm{e}}_s$ indicates the streamwise direction. In the present article, PCF is studied in both, non-tilted domains, viewed from the side in \textbf{(a)}, where $\hat{\bm{e}}_s$ is aligned with $x$, and in tilted domains, viewed from top in \textbf{(b)}, where $\hat{\bm{e}}_s$ is oriented at a non-zero angle $\theta$ relative to the $x$-dimension. For all values of $\theta$, the nondimensionalized laminar flow profile is $\bm{U}_0(y)=y \hat{\bm{e}}_s$. We visualize non-laminar PCF in the indicated midplane at $y=0$.}
\end{figure}

The velocity vector field $\bm{U}(x,y,z,t)$ and the pressure $p(x,y,z,t)$ in PCF are governed by the incompressible Navier-Stokes equations
\begin{align}
 \frac{\partial \bm{U}}{\partial t} + \left(\bm{U}\cdot \nabla \right)\bm{U} &=- \nabla p + \frac{1}{\mathrm{Re}}\nabla^2 \bm{U} \ ,\label{eq:nse1}\\
 \nabla \cdot \bm{U} &=0 \ ,\label{eq:nse2} 
\end{align}
in a three-dimensional channel. The channel domain is considered as periodic in the two lateral directions $x$ and $z$, such that the velocity field repeats in space as $\bm{U}(x,y,z,t)=\bm{U}(x+L_x,y,z,t)$ and $\bm{U}(x,y,z,t)=\bm{U}(x,y,z+L_z,t)$ with $L_x$ and $L_z$ the lateral domain sizes. The Navier-Stokes equations are nondimensionalized with the half-gap height $h$ and the wall velocity $U_w$ leading to the dimensionless Reynolds number $\mathrm{Re}=U_w h/\nu$. No-slip boundary conditions at the walls are imposed such that $\bm{U}(y=\pm1)=\pm \hat{\bm{e}}_s$. The unit vector $\hat{\bm{e}}_s=\cos(\theta_s)\hat{\bm{e}}_x + \sin(\theta_s)\hat{\bm{e}}_z$ describes the streamwise direction in which the walls move. The streamwise direction may be rotated by $\theta_s$ relative to the $x$- and $z$-direction of the rectangular domain. The nondimensionalized Navier-Stokes equations with these boundary conditions admit the linear velocity profile $\bm{U}_0(y)=y \hat{\bm{e}}_s$ as laminar flow solution (Fig.~\ref{fig:pcf:systemGeometry}).

We perform direct numerical simulations (DNS) of PCF in domains sufficiently large to contain turbulent-laminar stripes. Such large-domain DNS can be computationally demanding. We employ the MPI-parallel code CHANNELFLOW 2.0 \cite{Gibson2019}. The code implements a pseudo-spectral method based on Fouier-Chebychev decompositions of the velocity deviation field $\bm{u}(x,y,z,t)=\bm{U}(x,y,z,t)-\bm{U}_0(y)$. An implicit-explicit multistep algorithm of 3rd order is used for time marching. The mean pressure gradient is fixed at zero along the streamwise and the spanwise direction. We perform DNS of weakly turbulent PCF at $\mathrm{Re}=350$ in a large numerical domain of size $[L_x,L_z]=[197,87.5]$ with $[N_x,N_y,N_z]=[1024,33,512]$ spectral modes and walls moving with $\theta_s=0$ along the $x$-dimension. The flow self-organizes into oblique turbulent-laminar stripes predominantly oriented along the domain diagonal at $\theta=\pm 24^{\circ}$. The large-scale pattern is subject to statistical fluctuations in the pattern wavelength and orientation, as already observed previously \cite{Duguet2010}. Oblique stripe patterns may drift in space, break up, and form again. To reduce the complexity of the spatio-temporal dynamics, we will impose additional discrete symmetries of PCF. Imposing a discrete symmetry disallows instabilities that would spontaneously break this symmetry and reduces the number of degrees of freedom in the numerical simulation. Thereby, imposed symmetries also reduce the number of dimensions of the accessible discretized state space. The above described DNS-setup has $N_0=3\times (2N_x/3) \times N_y \times (2N_z/3) \approx 3.3\times 10^6$ degrees of freedom, corresponding to three velocity components on a de-aliased grid. In the following, the $N_0=3.3\times 10^6$ degrees of freedom are reduced by more than one order of magnitude. 

The governing equations (\ref{eq:nse1}-\ref{eq:nse2}), complemented by periodic boundary conditions and imposed wall velocity at $\theta_s=0^{\circ}$, are equivariant under reflections and translations in the $x$- and $z$-direction
\begin{align}
 \pi_{xy}[u,v,w](x,y,z,t) &=[-u,-v,w](-x,-y,z,t) \ , \label{eq:sympixy}\\
 \pi_{z}[u,v,w](x,y,z,t)&=[u,v,-w](x,y,-z,t) \ , \label{eq:sympiz}\\
 \tau(a_x,a_z)[u,v,w](x,y,z,t)&=[u,v,w](x+a_x L_x,y,z+a_z L_z,t) \label{eq:symtau}\ ,
\end{align}
with continuous real-valued shift factors $a_x,a_z \in [0,1)$. Symmetry transformations (\ref{eq:sympixy}-\ref{eq:symtau}) generate a symmetry group $S_{\mathrm{PCF}}=\langle \pi_{xy},\pi_z, \tau(a_x,a_z)\rangle$, where $\langle \rangle$ denotes all products between the listed transformations. For tilted domains with wall velocities at $\theta_s\neq 0$ and $-90^{\circ}<\theta_s <90$, the reflections $\pi_{xy}$ and $\pi_{z}$ are broken. Only their product $\pi_i=\pi_{xy}\pi_z$, the inversion symmetry
\begin{equation}
\pi_i[u,v,w](x,y,z,t) =[-u,-v,-w](-x,-y,-z,t) 
\end{equation}
remains a symmetry of PCF in the considered domain. To impose a particular symmetry $\sigma \in S_{\mathrm{PCF}}$ on a velocity field $\bm{u}$, we apply the projection $(\sigma \bm{u} + \bm{u})/2$. Such a projection requires the additional property $\sigma^2=1$. Once $\bm{u}$ is $\sigma$-symmetric, the time evolution of $\bm{u}$ will remain $\sigma$-symmetric because the governing equations are equivariant under $\sigma$ \citep{chaosbookSec10}. Thus, imposing symmetries on a flow confines the flow and its time evolution to a symmetry subspace containing only flows that are invariant under the imposed symmetries.

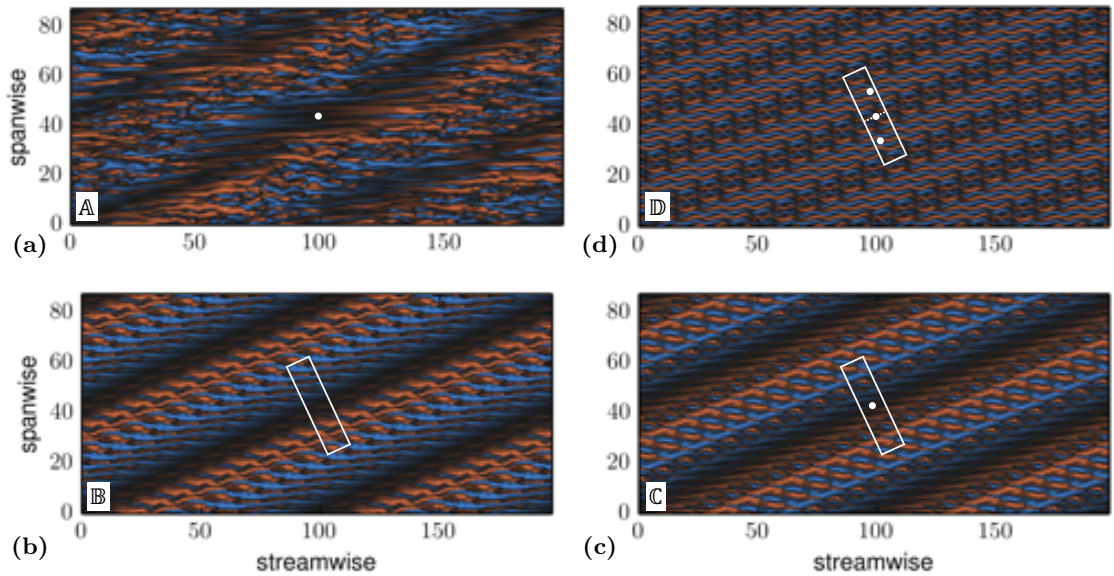
\begin{figure}[tb]
        \begin{tikzpicture}
    	\draw (0, 0) node[inner sep=0]{\includegraphics[width=0.99\linewidth,trim={0.0cm 0.2cm 0.0cm 0.2cm},clip]{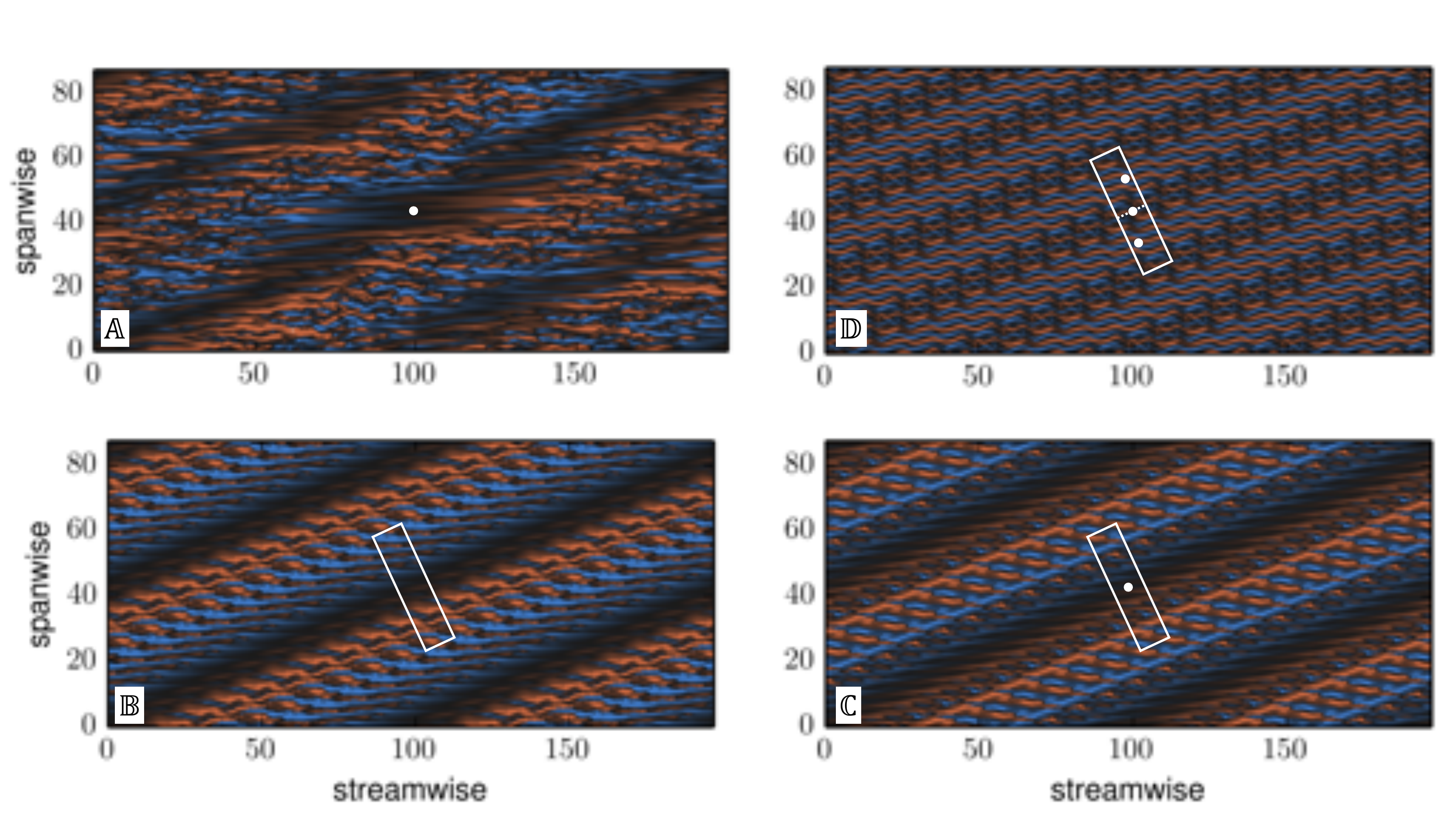}};
    	\draw (-7.0,0.3) node {\textbf{(a)}};
    	\draw (0.5,0.3) node {\textbf{(d)}};
    	\draw (-7.0,-3.7) node {\textbf{(b)}};
    	\draw (0.5,-3.7) node {\textbf{(c)}};
		\end{tikzpicture}
\caption{\label{fig:p4:4stripes} Imposing symmetries reduces the complexity of the oblique stripe pattern at $\mathrm{Re}=350$ from \textbf{(a)} to \textbf{(d)}. The contours indicate streamwise velocity at midplane ($y=0$). \textbf{(a)} Snapshot from a DNS in a periodic domain of extent $[L_x,L_z]=[197,87.5]$ with additionally enforced $\pi_i$-symmetry. \textbf{(b)} Periodically repeated snapshot from a DNS in a periodic domain of extent $[L_x,L_z]=[10,40]$ and tilted at $\theta=24^{\circ}$ against the streamwise direction. \textbf{(c)} Like in \textbf{(b)} but with inversion symmetry $\pi_i$ imposed additionally. \textbf{(d)} Like in \textbf{(c)} but with shift-inversion symmetry $\pi_{si}=\pi_i \tau(0.5,0.5)$ imposed additionally. This symmetry enforces a pattern wavelength of $\lambda=20$, while cases \textbf{(a-c)} have a pattern wavelength of $\lambda=40$. White lines outline the spatial extent of the periodic tilted domain. White dots mark the reference points for discrete inversion.}
\end{figure}

Changing the spatial extent of a double-periodic domain imposes discrete translation symmetries with different periods on the flow. Together with discrete symmetries, the periodic boundary conditions define the symmetry subspace of the flow. Previous numerical studies have systematically varied the domain size to investigate turbulent-laminar patterns in different symmetry subspaces \cite{Philip2011,Chantry2017}. Here, we aim for reducing the complexity of the spatio-temporal dynamics of a sustained oblique stripe pattern while trying to preserve the pattern characteristics of a large-scale wavelength and an oblique orientation. Starting from a regular stripe pattern in a large domain we systematically impose additional discrete symmetries in $S_{\mathrm{PCF}}$ or reduce the domain size. We consider simulations of oblique stripe patterns in four different symmetry subspaces, $\mathbb{A}$-$\mathbb{D}$, of PCF at $\mathrm{Re}=350$:

\begin{itemize}[leftmargin=0.7cm]
\item[ $\mathbb{A}$:] Inversion symmetry $\pi_i$ is imposed on PCF in the large periodic domain $[L_x,L_z]=[197,87.5]$ with $x$-aligned wall velocities, $\theta_s=0^{\circ}$. The number of degrees of freedom is $N_{\mathbb{A}}=N_0 /2=1\,686\,960$. A snapshot of a regular stripe pattern with pattern wavelength $\lambda=40$ and orientation at $\theta=24^{\circ}$ along the domain diagonal is shown in Fig.~\ref{fig:p4:4stripes}a. Imposing $\pi_i$ prohibits drift of the large-scale pattern by fixing the pattern's spatial phase in $x$ and $z$. In this case, the center of a laminar region coincides with the center of the domain (Fig.~\ref{fig:p4:4stripes}a).

\item[$\mathbb{B}$:] We impose the periodicity of a tilted periodic domain of extent $[L_x,L_z]=[10,40]$ and orientation $\theta_s=24^{\circ}$. The grid resolution is $[N_x,N_y,N_z]=[64,33,256]$. No reflection symmetry is imposed. The number of degrees of freedom is $N_{\mathbb{B}}=3\times (2N_x/3) \times N_y \times (2N_z/3) =706\,860$. The tilted domain allows to simulate a single spatial period of turbulent-laminar stripes whose geometry matches the boundary conditions of the domain with $\lambda=L_z=40$ and $\theta=\theta_s=24^{\circ}$. The domain is identical to the one used in \cite{Barkley2007}. A snapshot from the simulation is periodically repeated in Fig.~\ref{fig:p4:4stripes}b.

\item[$\mathbb{C}$:] Again, a single stripe period is simulated in a tilted domain like for subspace $\mathbb{B}$ but with additionally imposed $\pi_i$-symmetry. The corresponding symmetry subspace, resolved with $N_{\mathbb{C}}=N_{\mathbb{B}} /2=353\,430$ degrees of freedom, contains the mean flow of the stripe pattern in a tilted domain of size $[L_x,L_z]=[10,40]$ \cite{Barkley2007} and also the stripe equilibrium reported in \cite{Reetz2019a}. The emerging oblique stripe pattern is shown in Fig.~\ref{fig:p4:4stripes}c.

\item[$\mathbb{D}$:] In the final step of reducing the complexity of the dynamics, we impose a shift-inversion symmetry $\pi_{si}=\pi_i \tau(0.5,0.5)$ in addition to the symmetries of subspace $\mathbb{C}$. The number of degrees of freedom is $N_{\mathbb{D}}=N_{\mathbb{B}} /4=176\,715$. The $\pi_{si}$-symmetry changes the wavelength of the emerging oblique stripe pattern from $\lambda=40$ to $\lambda=20$. Oblique stripe patterns with $\pi_{si}$-symmetry and wavelength $\lambda=20$ have been observed in the bifurcation sequence towards the stripe equilibrium solution \cite{Reetz2019a} but are typically not naturally selected in less confined domains.
\end{itemize}

\begin{figure}[tb]
        \begin{tikzpicture}
    	\draw (0, 0) node[inner sep=0]{\includegraphics[width=0.99\linewidth,trim={1.0cm 0.0cm 0.5cm 0.0cm},clip]{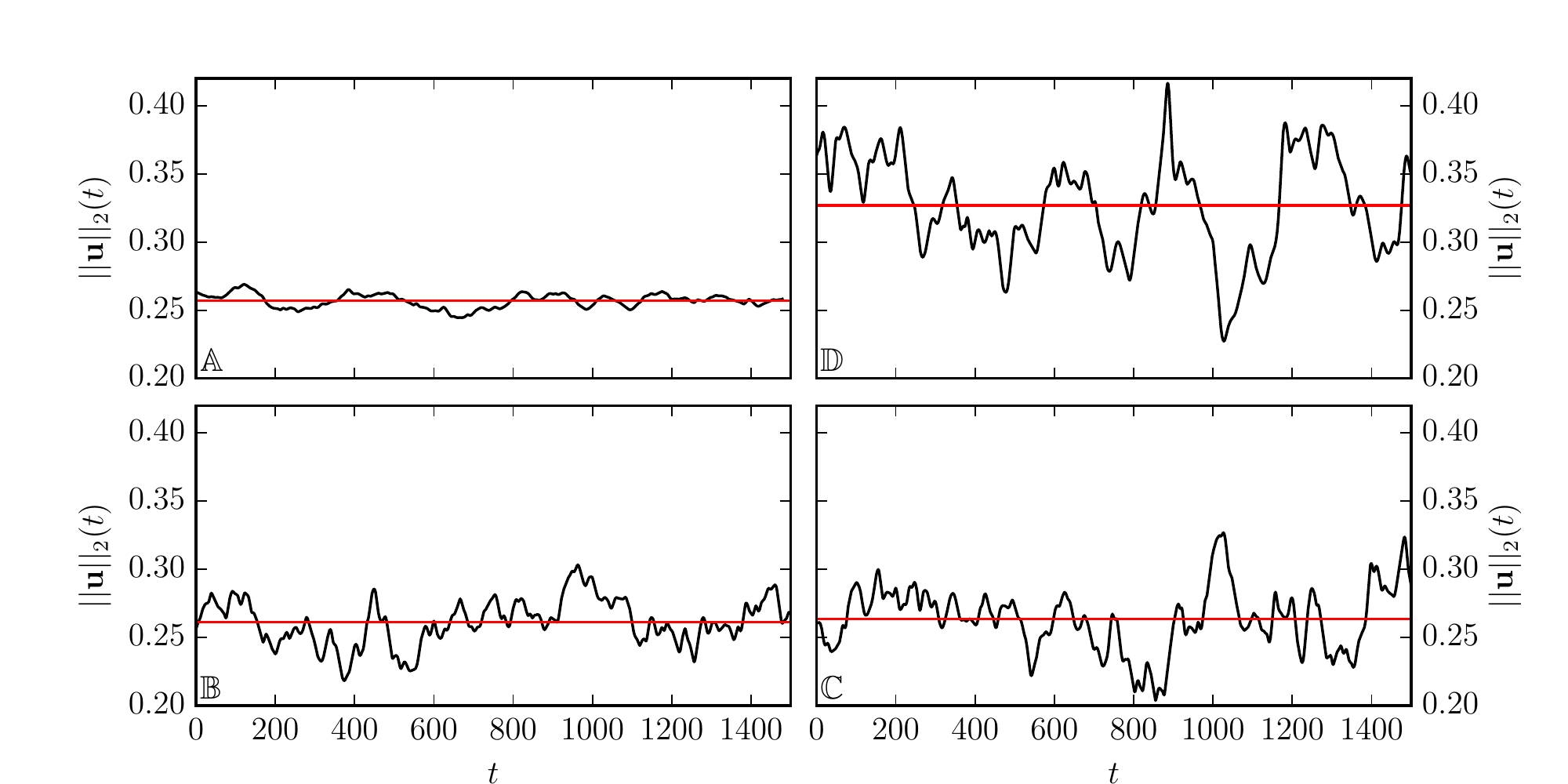}};
    	\draw (-7.3,0.2) node {\textbf{(a)}};
    	\draw (7.2,0.2) node {\textbf{(d)}};
    	\draw (-7.3,-3.7) node {\textbf{(b)}};
    	\draw (7.2,-3.7) node {\textbf{(c)}};
		\end{tikzpicture}
\caption{\label{fig:p4:4timeseries} Time series from DNS in four different symmetry subspaces corresponding to the four snapshots in Fig.~\ref{fig:p4:4stripes} \textbf{(a-d)} showing instance $t=0$ in \textbf{(a-d)} of the present figure. While the amplitudes of the temporal fluctuations differs significantly between the cases \textbf{(a-d)}, the mean values of $||\bm{u}||_2$ (horizontal red lines) is approximately equal in the symmetry subspaces of cases \textbf{(a-c)}. Case \textbf{(d)} has a different mean because the imposed shift-inversion symmetry disallows the oblique stripe patterns at wavelength $\lambda=40^{\circ}$, present in \textbf{(a-c)}, and enforces a pattern wavelength of  $\lambda=20$. }
\end{figure}
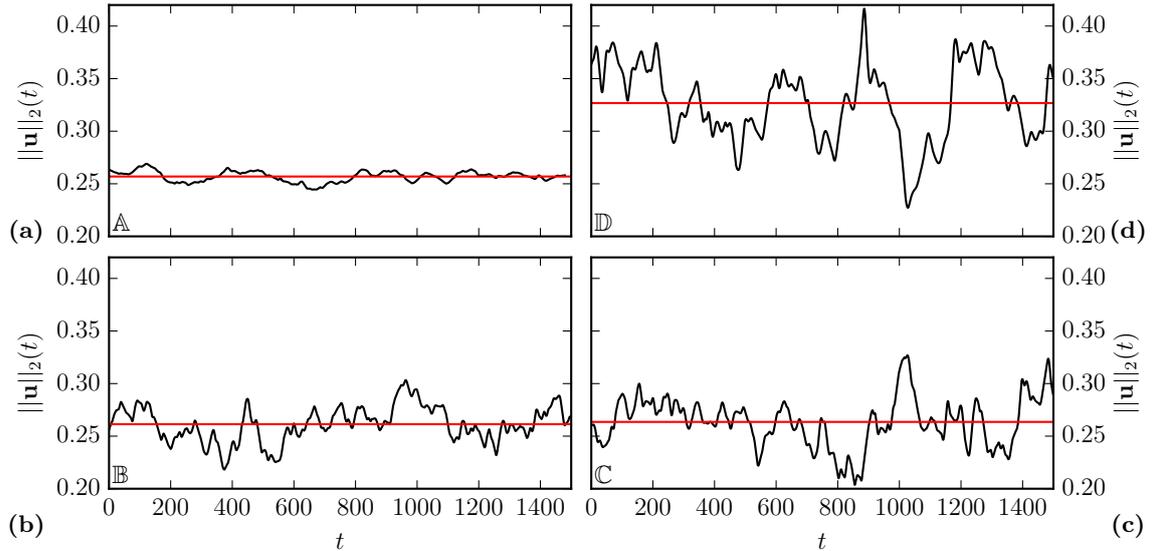

Robust oblique stripe patterns are observed in all four DNS runs over a time interval of $\Delta t = 1\,500$.  Beyond this time interval, stripes typically either break up, leading to defects in the large-scale pattern (observed for DNS in $\mathbb{A}$), or decay to laminar flow (observed for DNS in $\mathbb{B}$-$\mathbb{D}$). In the time interval over which the pattern characteristics are robust, the domain averaged velocity norm 
\begin{equation}
||\bm{u}||_2(t)=\frac{1}{(2L_x L_z)^{1/2}} \left( \int_0^{L_z}\int_{-1}^{1}\int_0^{L_x} \bm{u}^2(x,y,z,t)\,\mathrm{d}x\,\mathrm{d}y\,\mathrm{d}z \right)^{1/2}
\end{equation}
shows chaotic oscillations around its mean of different oscillation amplitude (Fig.~\ref{fig:p4:4timeseries}). In the higher-dimensional subspace $\mathbb{A}$, temporal velocity fluctuations at different uncorrelated locations in the domain statistically compensate each other more than in the lower-dimensional subspace $\mathbb{D}$. There, the imposed symmetries lead to spatial correlations that reduce the number of independently fluctuating modes. This induces larger fluctuations. Thus, the fluctuation amplitude of norm $||\bm{u}||_2(t)$ around its mean is a proxy for the spatio-temporal complexity of the pattern dynamics at equal $\mathrm{Re}$.\\
The temporal average of $||\bm{u}||_2(t)$ over $\Delta t = 1500$ is approximately identical for oblique stripes in subspaces $\mathbb{A}$-$\mathbb{C}$ (red lines in Fig.~\ref{fig:p4:4timeseries}), suggesting that the imposed symmetries do not change the pattern's mean flow. In subspace $\mathbb{D}$ however, the shift-inversion symmetry $\pi_{si}$ enforces a pattern wavelength of $\lambda=20$. This increases the temporal mean from $||\bm{u}||_2=0.26$, observed for stripes with $\lambda=40$, to $||\bm{u}||_2=0.33$. Thus, the stripe pattern in symmetry subspace $\mathbb{D}$ is qualitatively and quantitatively different from the stripe patterns in $\mathbb{A}$-$\mathbb{C}$. All stripe patterns are however obliquely oriented at $\theta=24^{\circ}$ and periodic at wavelength $\lambda=40$. Note that subspace $\mathbb{A}$ is chosen to accommodate a pattern of this orientation angle and wavelength but does not contain the subspaces $\mathbb{B}$-$\mathbb{D}$. Subspace $\mathbb{B}$ however contains $\mathbb{C}$ and $\mathbb{D}$. Including the subspaces $\mathbb{B}$-$\mathbb{D}$ of the tilted domain into the subspace of a non-tilted domain requires choosing a non-tilted domain with at least $[L_x,L_z]=[98.5,394]$. The geometric condition to make the flow in tilted and non-tilted domains commensurable is discussed in \cite{Reetz2019a}. There is no obvious additional symmetry to further reduce the complexity of the spatio-temporal dynamics of oblique stripe patterns beyond subspace $\mathbb{D}$. 
 
\section{Edge of chaos in symmetry subspaces}\label{sec:p4:edge}

\begin{figure}[tb]
        \begin{tikzpicture}
    	\draw (0, 0) node[inner sep=0]{\includegraphics[width=0.99\linewidth,trim={1.0cm 4.2cm 0.5cm 0.0cm},clip]{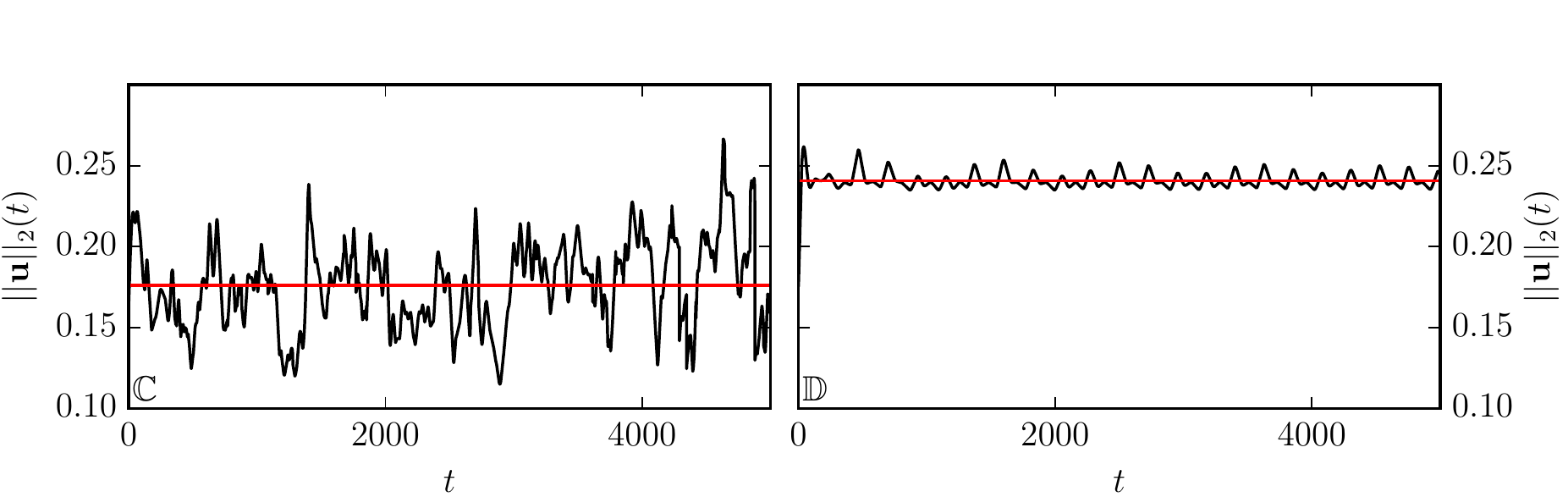}};
    	\draw (-7.3,-2.0) node {\textbf{(a)}};
    	\draw (7.2,-2.0) node {\textbf{(b)}};
		\end{tikzpicture}
\caption{\label{fig:p4:edgetracking} Time series of edge tracking at $\mathrm{Re}=350$ in different symmetry subspaces. Both results are computed in a periodic domain of extent $[L_x,L_z]=[10,40]$ and tilted at $\theta=24^{\circ}$ against the streamwise direction. In addition, centro-inversion $\pi_i$ is imposed in \textbf{(a)}, and shift-inversion $\pi_{si}$ is imposed in \textbf{(b)}. The mean values of $||\bm{u}||_2$ (horizontal red lines) differ like in DNS in the same two symmetry subspaces (Fig.~\ref{fig:p4:4timeseries}c,d). Only when imposing $\pi_{si}$ the edge-tracking approach yields a simple edge state. }
\end{figure}

Despite reducing the number of degrees of freedom to the presumable minimum for supporting oblique stripe patterns, the dynamics in the symmetry subspaces remains chaotic. To identify periodic orbits, we follow the established approach to confine the dynamics to the edge of chaos. Using the edge-tracking algorithm implemented in CHANNELFLOW 2.0 \cite{Gibson2019}, we follow two trajectories inside the edge of chaos in symmetry subspace $\mathbb{C}$ and $\mathbb{D}$, respectively. The Reynolds number is again fixed at $\mathrm{Re}=350$. The initial condition was chosen arbitrarily from the DNS in $\mathbb{C}$ and $\mathbb{D}$. We confirmed that the state approached by edge-tracking does not depend on the initial condition in these two cases. Edge-tracking in $\mathbb{C}$ follows a chaotic trajectory indicating a chaotic edge state (Fig.~\ref{fig:p4:edgetracking}a). Chaotic edge states have been described previously for pipe flow \cite{Schneider2007b} and PCF \cite{Duguet2009}. In $\mathbb{D}$, the trajectory approaches a near-periodic edge state (Fig.~\ref{fig:p4:edgetracking}b). 

The chaotic edge state in $\mathbb{C}$ differs clearly from the chaotic state found by DNS in $\mathbb{C}$ (Fig.~\ref{fig:p4:4timeseries}c). The trajectory in the edge of chaos in $\mathbb{C}$ has a mean $L_2$-norm of $||\bm{u}||_2=0.18$ which is significantly smaller than the mean of the trajectory in the DNS, $||\bm{u}||_2=0.26$. Likewise, the near-periodic edge state in $\mathbb{D}$ has an $L_2$-norm of $||\bm{u}||_2=0.24$, lower than the simulated state in $\mathbb{D}$ with $||\bm{u}||_2=0.33$ (Fig.~\ref{fig:p4:4timeseries}d). 

In summary, a chaotic edge state is found in $\mathbb{C}$. By additionally imposing the shift-inversion symmetry $\pi_{si}$, a much simpler, near-periodic edge state is found in $\mathbb{D}$.

\section{Unstable periodic orbits}\label{sec:p4:orbits}

The near-periodic oscillations in the edge of chaos in symmetry subspace $\mathbb{D}$ suggest the presence of an unstable periodic orbit (UPO). Periodic orbits satisfy the recurrency condition
\begin{equation} \label{eq:p8:map}
 \mathcal{F}^T(\bm{u}(x,y,z,t);\mathrm{Re})-\bm{u}(x,y,z,t)=0 \ .
\end{equation}
The operator $\mathcal{F}^T(\bm{u};\mathrm{Re})$ evolves PCF according to the Navier-Stokes equations at a specific $\mathrm{Re}$ from the initial velocity field $\bm{u}$ over time period $T$. A velocity field solving (\ref{eq:p8:map}) can be found via Newton-Raphson iteration. Equation (\ref{eq:p8:map}) is solved with a matrix-free Newton-Krylov method. We use the Newton-Krylov method implemented in the nonlinear solver library of CHANNELFLOW 2.0 \cite{Gibson2019} to converge and numerically continue invariant solutions under changing $\mathrm{Re}$ or domain size. In order to converge the periodic orbits discussed here, a multi-shooting method with two shots is required \cite{Gibson2019}.

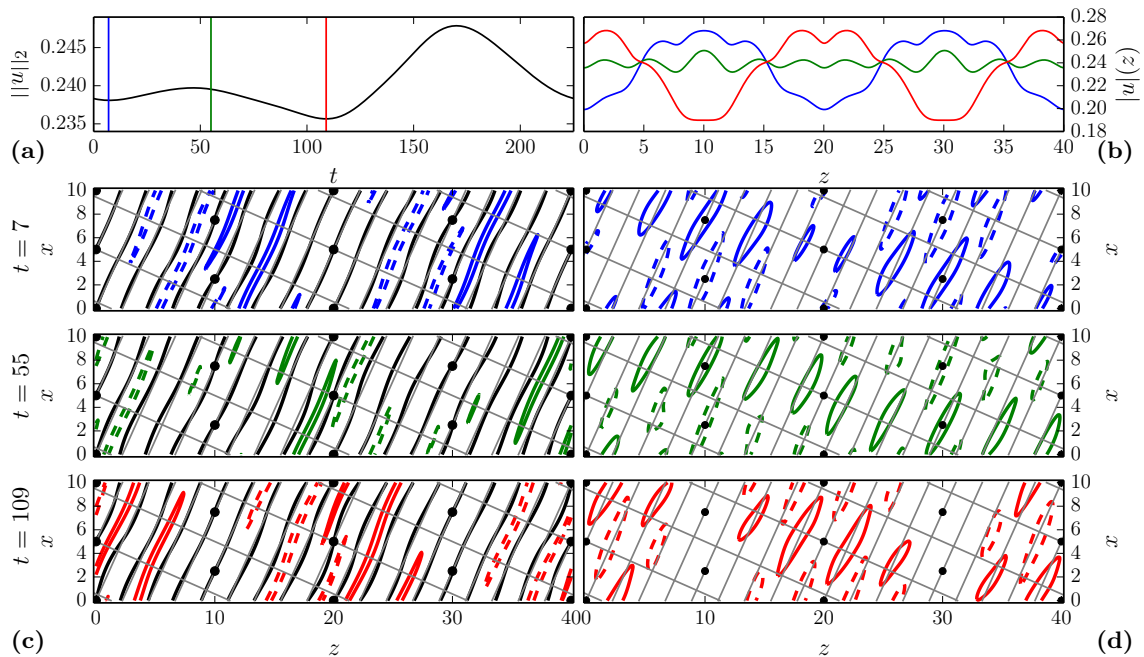
\begin{figure}[tb]
        \begin{tikzpicture}
    	\draw (0, 0) node[inner sep=0]{\includegraphics[width=0.99\linewidth,trim={1.5cm 0.5cm 1.0cm 0.0cm},clip]{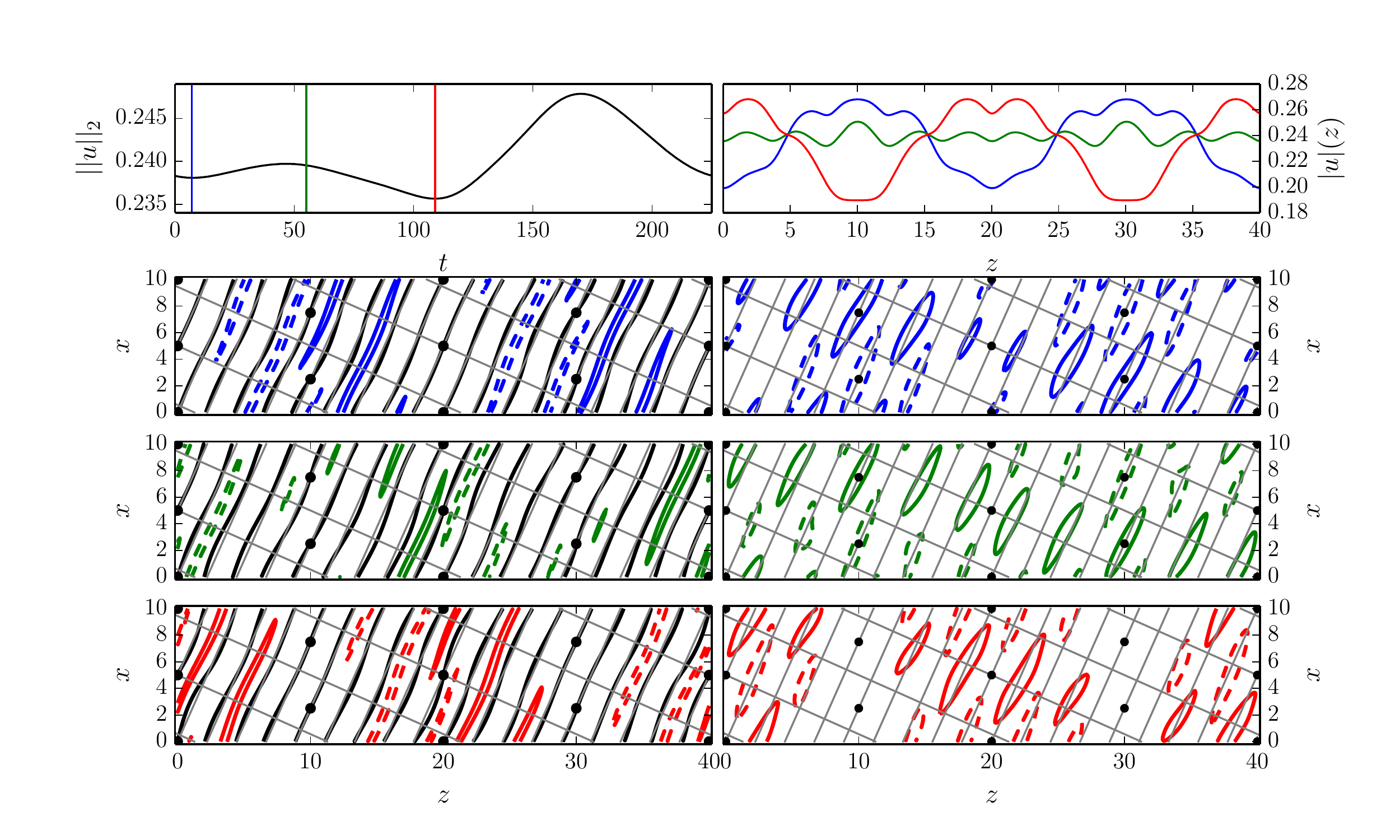}};
    	\draw (-7.2,2.0) node {\textbf{(a)}};
    	\draw (7.1,2.0) node {\textbf{(b)}};
    \draw (-7.2,-4.5) node {\textbf{(c)}};
    	\draw (7.1,-4.5) node {\textbf{(d)}};
    	\draw (-7.3,0.75) node [rotate=90]{$t=7$};
    	\draw (-7.3,-1.1) node [rotate=90]{$t=55$};
    	\draw (-7.3,-2.95) node [rotate=90]{$t=109$};
		\end{tikzpicture}
\caption{\label{fig:p4:stripeOrbit} Unstable periodic orbit $OSW_1$ in the edge of chaos in subspace $\mathbb{D}$ describes a standing wave modulation. \textbf{(a)} Velocity norm $||\bm{u}||_2$ oscillations over period $T=225.4$ with two local maxima and minima. The local minima correspond to the instances of large amplitude modulations as indicated by the $x$-$y$-averaged root mean square velocity modulation $|u|(z)=(2L_x)^{-1/2} (\iint \bm{u}^2\,\mathrm{d}x\mathrm{d}y)^{1/2}$ in \textbf{(b)}. These instances are further illustrated by midplane contours of streamwise velocity \textbf{(c)} (solid at $\bm{u}_s=0.45$, dashed at $\bm{u}_s=-0.45$), and midplane contours of streamwise vorticity \textbf{(d)} (solid at $\bm{\omega}_s=0.11$, dashed at $\bm{\omega}_s=-0.11$). The grey grid indicates the streamwise and the spanwise directions, tilted at $\theta=24^{\circ}$ against the domain dimensions. The black contours in \textbf{(c)} mark the critical layer with streamwise velocity $u_s=0$. Locally in space, the UPO oscillates between high-amplitude wavy velocity streaks generating much vorticity and low-amplitude near-straight velocity streaks generating little vorticity. }
\end{figure}

Using a velocity field obtained from edge-tracking in subspace $\mathbb{D}$ at $\mathrm{Re}=350$ as initial state for the Newton-Krylov iteration yields a UPO with period $T=225.4$. The UPO is composed of wavy velocity streaks whose amplitude is modulated in space and oscillates in time along the orbit. To clearly distinguish signals in time and space, we use `modulation' for spatial signals and `oscillation' for temporal signals. The oscillating amplitude modulation represents a standing wave with a dominant pattern wavelength of $\lambda=20$ along the $z$-direction of the tilted domain (Fig. \ref{fig:p4:stripeOrbit}). The standing wave has anti-nodes at $z\in \{0,10,20,30,40\}$ the points about which inversion symmetries are imposed. The locations of these symmetry points are marked by black dots in midplane sections showing velocity and vorticity contours in Fig. \ref{fig:p4:stripeOrbit}c,d. Instances $t=7$ and $t=109$ along the orbit show spatially localized regions of high velocity and vorticity fluctuations. These localized regions are centered around the symmetry points (see blue and red contours in Fig. \ref{fig:p4:stripeOrbit}c,d). Regions of high velocity and vorticity fluctuations have also a large $x$-$y$-averaged velocity norm (Fig. \ref{fig:p4:stripeOrbit}b) and coincide with velocity streaks that are more wavy than outside these regions where streaks are more straight (Fig. \ref{fig:p4:stripeOrbit}c). Thus, the UPO is a standing wave that along a temporal cycle periodically exchanges regions of high amplitude wavy velocity streaks and low amplitude near-straight velocity streaks. The evolving oblique amplitude modulation represents a large-scale pattern at half the wavelength of previously studied oblique stripes \cite{Reetz2019a}, as visualized in the supplementary video highlighting the UPO's dynamics \cite{supplement}. We name this UPO ``oblique standing wave" and denote it by $OSW_1$.

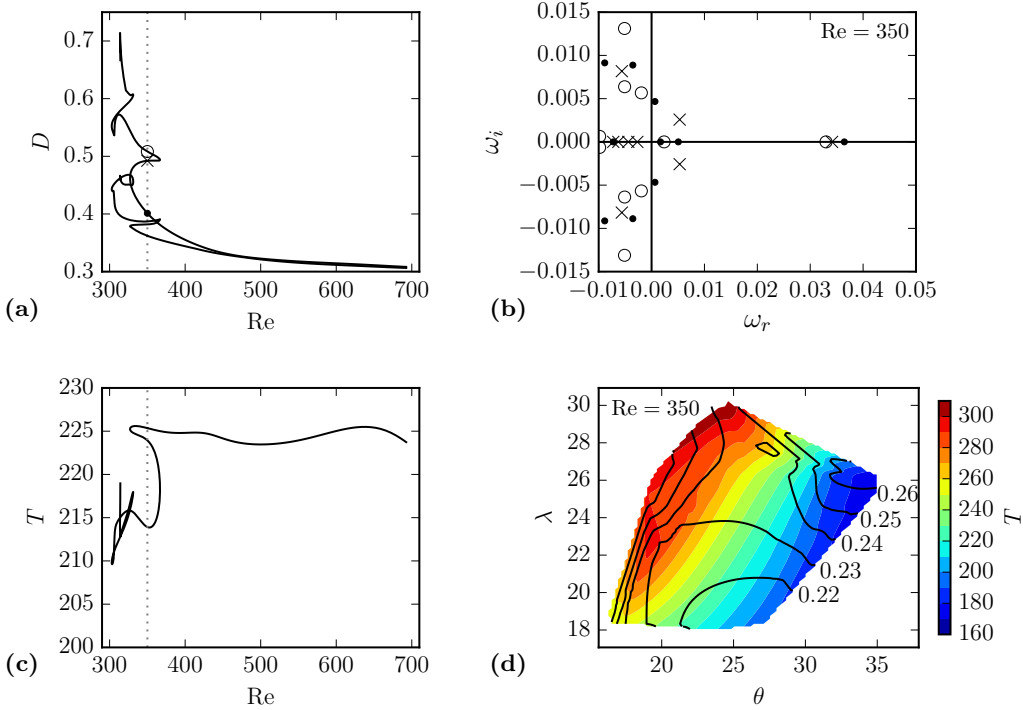
\begin{figure}[tb]
\center
        \begin{tikzpicture}
    	\draw (0, 0) node[inner sep=0]{\includegraphics[width=0.95\linewidth,trim={0.3cm 1.3cm 1.0cm 1.0cm},clip]{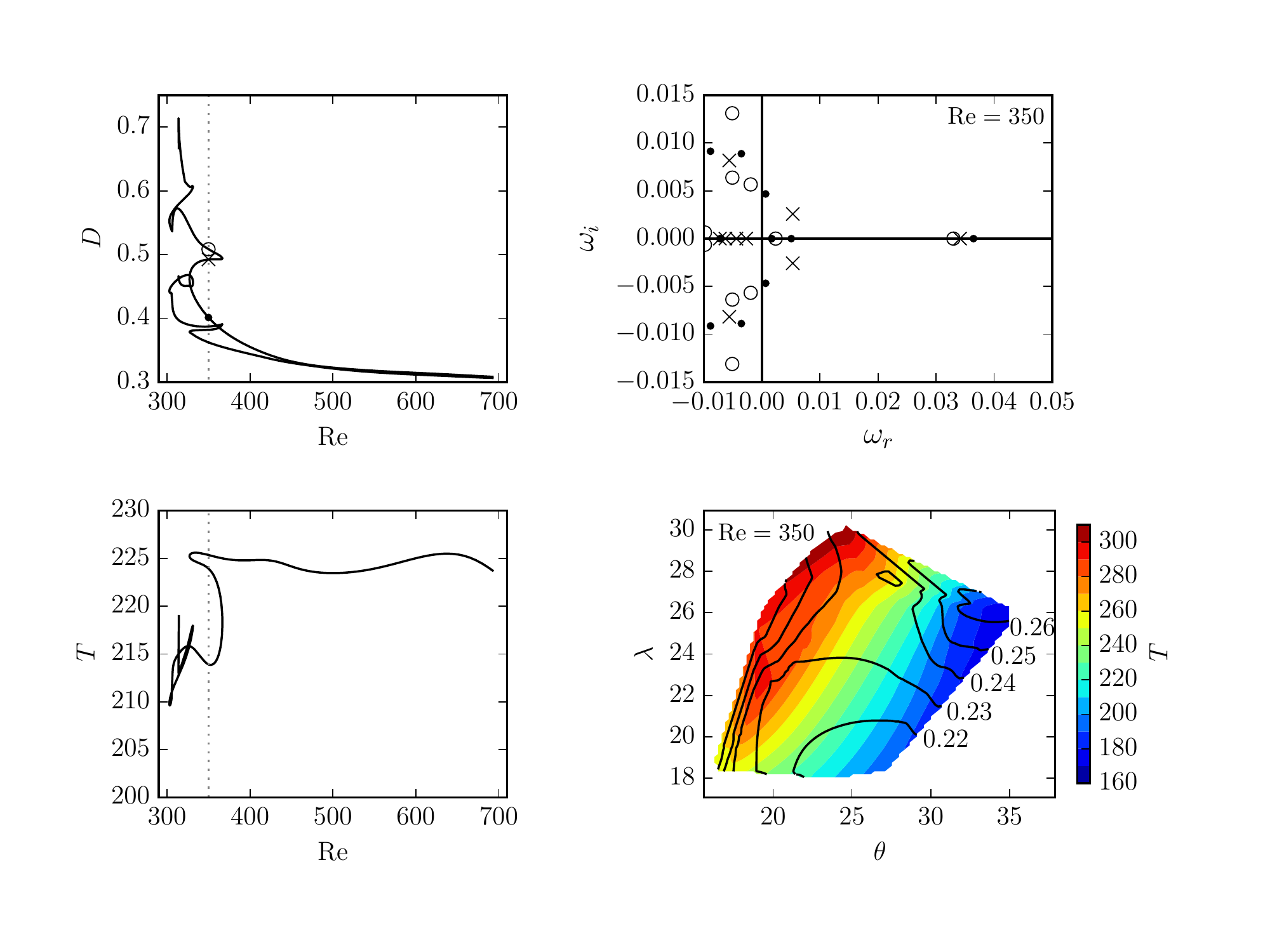}};
    	\draw (-6.5,0.5) node {\textbf{(a)}};
    	\draw (-0.1,0.5) node {\textbf{(b)}};
    	\draw (-6.5,-4.2) node {\textbf{(c)}};
    	\draw (-0.1,-4.2) node {\textbf{(d)}};
		\end{tikzpicture}
\caption{\label{fig:p4:bifDiag}  \textbf{(a)} Bifurcation diagram of the unstable periodic orbit $OSW_1$ along $\mathrm{Re}$, computed via numerical continuation. The two curves indicate maximum and minimum dissipation $D$ over one period. Towards high $\mathrm{Re}$, the oscillation amplitude reduces. Towards low $\mathrm{Re}$, the branch undergoes folds leading to three periodic orbits, $OSW_{1,2,3}$, coexisting at $\mathrm{Re}=350$ (dashed line). \textbf{(b)} Spectrum of eigenvalues of the three unstable periodic orbits at $\mathrm{Re}=350$: $OSW_1$ ($\bullet$), $OSW_2$ ($\times$), $OSW_3$ ($\bigcirc$). The spectrum is calculated in the symmetry subspace $\mathbb{D}$. \textbf{(c)} Bifurcation diagram as in \textbf{(a)} but indicating period $T$ of the continued periodic orbits. \textbf{(d)} Existence range of $OSW_1$ at $\mathrm{Re}=350$ under changing orientation angle $\theta$ and pattern wavelength $\lambda$. Existence boundaries coincide with fold bifurcations. Colored contours indicate period $T$. Black contour lines with numbers indicate phase velocity $c$ defined in (\ref{eq:p4:phasevelocity}).  }
\end{figure}

When $OSW_1$ is numerically continued up in $\mathrm{Re}$, the amplitude at which the orbit's total dissipation $D$ oscillates in time reduces. The reduced oscillation amplitude remains approximately constant for $450<\mathrm{Re}<700$ (Fig. \ref{fig:p4:bifDiag}a). For decreasing $\mathrm{Re}$, the oscillation amplitude increases and the solution branch undergoes a sequence of two smooth folds at $\mathrm{Re}=327$ and $\mathrm{Re}=367$, respectively. Further along the branch a succession of sharp folds emerges in the interval $300<\mathrm{Re}<330$ (Fig. \ref{fig:p4:bifDiag}a). We have observed such sharp and irregular folds previously along branches of equilibrium solutions underlying stripes \cite{Reetz2019a}. The folded solution branches connect three periodic orbits at $\mathrm{Re}=350$ (Fig.~\ref{fig:p4:bifDiag}a). We index these orbits according to the order at which they are encountered along the branch, $OSW_{i}$ with $i=1,2,3$.

The period $T$ of $OSW_1$ shows little variation when $\mathrm{Re}$ is changed. The largest variation in the period occurs across the second fold from $OSW_2$ to $OSW_3$ where the period drops from $T=225$ to $T=215$ (Fig.~\ref{fig:p4:bifDiag}c). In contrast to changes in $\mathrm{Re}$, the period of $OSW_1$ depends significantly on changes in the pattern wavelength $\lambda$ and the orientation angle $\theta$ (Fig.~\ref{fig:p4:bifDiag}d). In a tilted minimal domain, the boundary conditions control $\lambda$ and $\theta$. For $\lambda$-continuation at constant $\theta$, we vary the domain diagonal at fixed aspect ratio $L_x/L_z$. For $\theta$-continuation at constant $\lambda$, we adjust $L_x$ according to $L_x(\theta)\sim L_z/\tan(\theta)$ at fixed $L_z$. Orbit $OSW_1$ is continued along $\lambda$ and $\theta$ until fold bifurcations are encountered along the branches. These folds mark the limits of existence of $OSW_1$. For $\mathrm{Re}=350$, the orbit $OSW_1$ exists within the intervals $18<\lambda <30$ and $16^{\circ}<\theta <35^{\circ}$ (Fig.~\ref{fig:p4:bifDiag}d).

The periodic orbit $OSW_1$ exhibits a standing wave modulation. The standing wave can be decomposed into two counter-propagating traveling waves with a streamwise phase velocity of absolute value
\begin{equation}\label{eq:p4:phasevelocity}
 c= \frac{\lambda}{ T\,\sin(\theta)} \ .
\end{equation}
Based on the numerical continuation of $OSW_1$ in both, the pattern wavelength $\lambda$ and the orientation angle $\theta$, the functional form $c(\lambda,\theta)$ is calculated and visualized in Fig.~\ref{fig:p4:bifDiag}d. Despite the significant changes in $\lambda$, $\theta$ and period $T$, the combined ratio (\ref{eq:p4:phasevelocity}) varies only in the interval $0.21<c<0.27$. The approximately constant phase velocity $c$ suggests a scaling relation. Hence, we can use linear extrapolation to predict the approximate period of oblique wave solutions that are analogous to $OSW_1$ but have a pattern wavelength of $\lambda=40$. A wavelength of $\lambda=40$ is typically observed for self-organized oblique stripes in less constrained domains. Linear extrapolation of $T$ along $\lambda$ at $\theta =24^{\circ}$ predicts a period of $T\approx 380$ for oblique standing waves with $\lambda=40$. Note that the discussed oblique standing waves have slow temporal dynamics with periods on the order of the viscous diffusion time scale, $T\sim h^2/\nu=350$.

The dynamical stability in subspace $\mathbb{D}$ of each orbit is characterized by calculating the spectrum of eigenvalues at $\mathrm{Re}=350$ using Arnoldi iteration. All three periodic orbits are dynamically unstable and have one dominating purely real unstable eigenvalue of $\omega_r\approx 0.035$. Moreover, all three periodic orbits have at least one additional unstable eigenvalue in the interval $0<\omega_r<0.01$ (Fig.~\ref{fig:p4:bifDiag}b). Consequently, none of the three periodic orbits $OSW_{1,2,3}$ is an edge state which requires a single unstable eigenvalue \cite{Skufca2006,Schneider2007b}. Specifically, $OSW_{1}$ has in total five unstable eigenvalues. Three of them are real and the remaining two form a complex pair. $OSW_{2}$ has in total three unstable eigenvalues. One of them is real and the remaining two form a complex conjugate pair. $OSW_{3}$ has two purely real unstable eigenvalues and thus, has the lowest-dimensional unstable eigenspace of all three periodic orbits.

\begin{figure}[tb]
        \begin{tikzpicture}
    	\draw (0, 0) node[inner sep=0]{\includegraphics[width=0.99\linewidth,trim={0.9cm 1.1cm 0.1cm 0.0cm},clip]{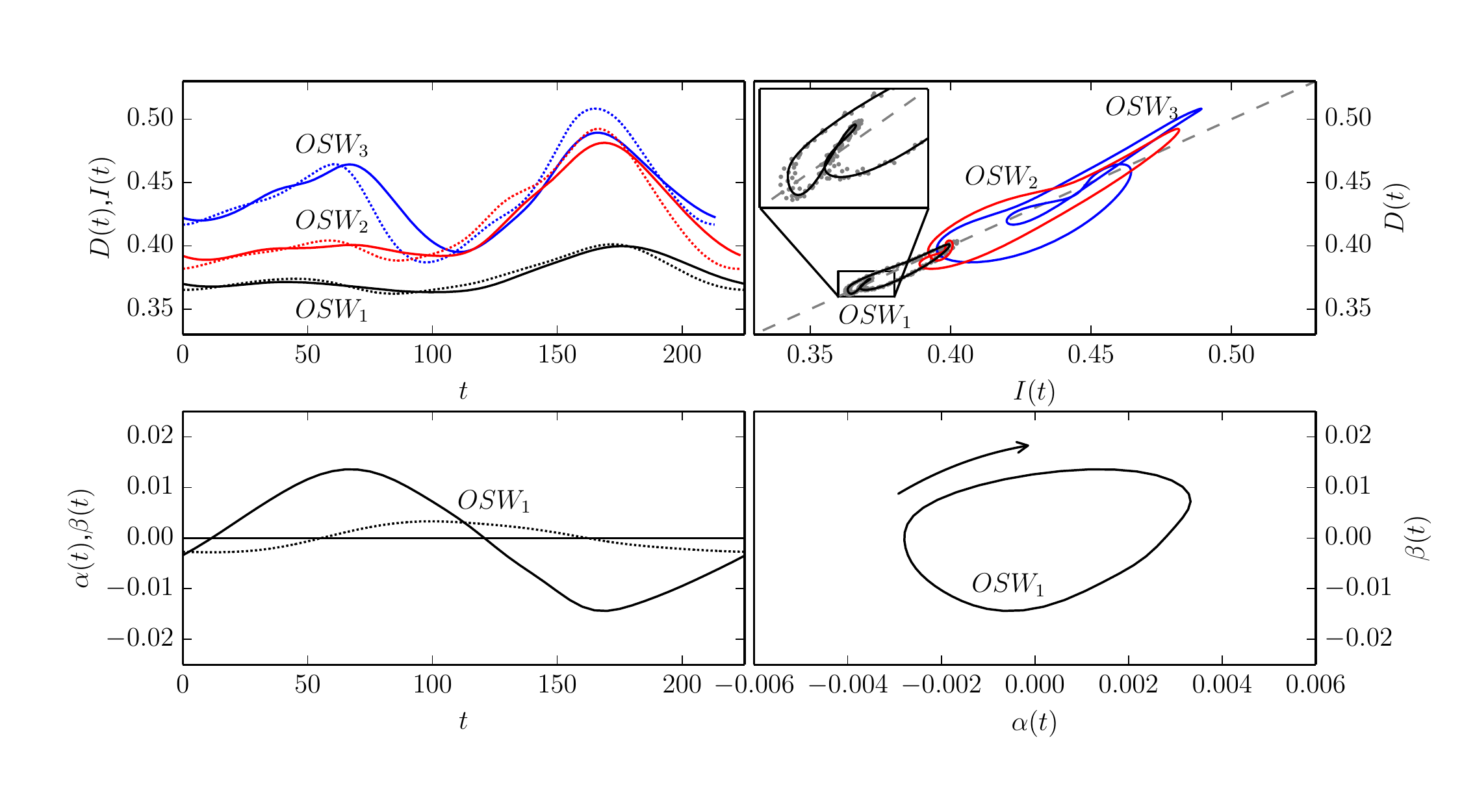}};
    	\draw (-7.0,0.3) node {\textbf{(a)}};
    	\draw (6.7,0.3) node {\textbf{(b)}};
    \draw (-7.0,-3.4) node {\textbf{(c)}};
    	\draw (6.7,-3.4) node {\textbf{(d)}};
		\end{tikzpicture}
\caption{\label{fig:p4:edgeOrbit} Two different projections for a phase portrait of the three unstable periodic orbits $OSW_{1,2,3}$ underlying slow oblique standing waves. Time series of kinetic energy input $I(t)$ and dissipation $D(t)$, defined in (\ref{eq:p4:input}) and (\ref{eq:p4:dissipation}), and of $\alpha(t)$ and $\beta(t)$, defined in (\ref{eq:p4:alpha}) and (\ref{eq:p4:beta}) are plotted in in panels \textbf{(a)} and \textbf{(c)}, respectively. The corresponding phase portraits are plotted in \textbf{(b)} and \textbf{(d)}. In addition to the trajectories of $OSW_{1,2,3}$, the edge-tracking trajectory for $9000<t<10000$ in subspace $\mathbb{D}$ is shown (grey dots in panel b) and clearly indicates orbit $OSW_{1}$ as part of the attractor in the edge of chaos. However, the distribution of points along this trajectory does not exactly coincide with the trajectory of $OSW_{1}$ (inset in panel b). This suggests that $OSW_{1}$ is part of a more complex attractor in the edge of chaos.  }
\end{figure}
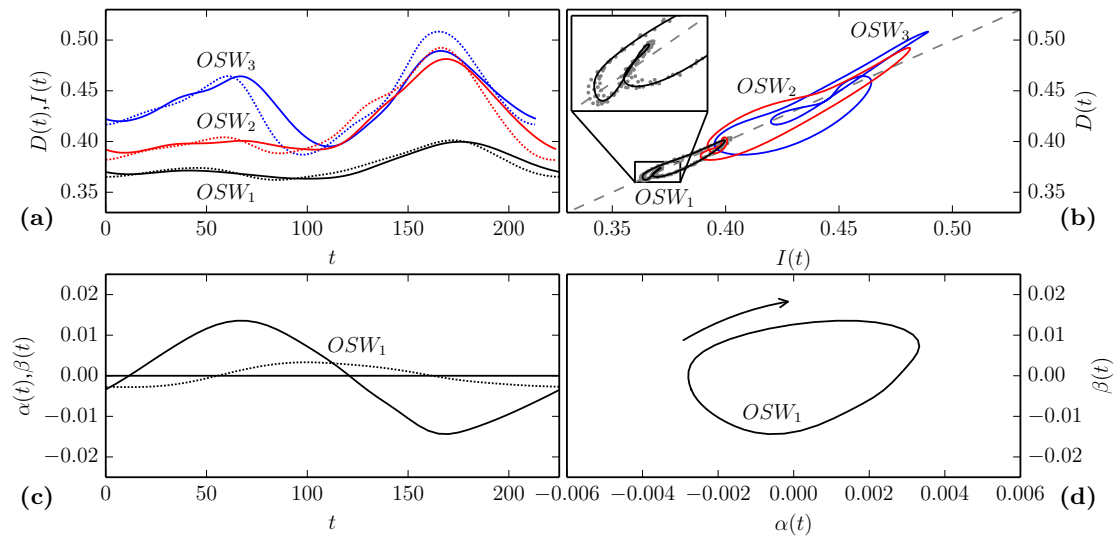

None of the three UPOs is an edge state. This raises the question if all of them are part of the attractor in the edge of chaos. To study the attractor in the edge of chaos, edge-tracking is performed for an additional $10\,000$ advective time units $h/U_w$. The state space trajectory over the last $1\,000$ time units is projected onto a plane indicating kinetic energy input $I(t)$ and dissipation $D(t)$, where
\begin{align}
 I(t)&=1+\frac{1}{2A} \int_A \left( \frac{\partial u_s}{\partial y}\bigg|_{y=-1}+ \frac{\partial u_s}{\partial y}\bigg|_{y=1} \right)\mathrm{d}A \label{eq:p4:input}\\
 D(t)&=\frac{1}{V}\int_{\Omega} | \nabla\times(\bm{u}+y\hat{\bm{e}}_s)|^2 \mathrm{d}\Omega \label{eq:p4:dissipation}
\end{align}
with the streamwise unit vector $\hat{\bm{e}}_s=\cos(\theta_s)\hat{\bm{e}}_x + \sin(\theta_s)\hat{\bm{e}}_z$ and the streamwise velocity component $u_s=\bm{u}\cdot \hat{\bm{e}}_s$. The quantities are normalized by cross-sectional area $A=L_xL_z$ and volume $V=2L_xL_z$ of the numerical domain, respectively. In addition to the edge trajectory we show the state space trajectories of all orbits $OSW_{1,2,3}$. The projection yields a phase portrait (Fig.~\ref{fig:p4:edgeOrbit}b) that clearly reveals how the edge-tracking trajectory (grey dots) clusters around $OSW_1$ but not around $OSW_2$ or $OSW_3$. Thus, of the three UPOs only $OSW_1$ is part of the attractor in the edge of chaos. The edge-tracking trajectory, even after $9\,000$ time units, does not coincide with $OSW_1$ (Inset in Fig. \ref{fig:p4:edgeOrbit}b). This is in line with the above observation that $OSW_1$ is not an edge state and may only be part of an attractor in the edge of chaos.

For each orbit, its time series for $I(t)$ and $D(t)$ almost coincide along the orbits (Fig.~\ref{fig:p4:edgeOrbit}a). As a consequence of the highly correlated $I(t)$ and $D(t)$, the phase portrait in Fig.~\ref{fig:p4:edgeOrbit}b shows orbits that are entangled and elongated along the diagonal line $D=I$. The highly correlated quantities $I(t)$ and $D(t)$ do not provide a good projection to illustrate the oscillatory behavior along the three orbits. A second projection is defined in terms of the two quantities 
\begin{align}
 \alpha(t)&= \mathbb{R}\left\{ \widetilde{\bm{u}^2}_{i=0,j=0,k=2} \right\}  \ ,  \label{eq:p4:alpha}\\
 \beta(t)&=  \mathbb{R}\left\{ \widetilde{\bm{\omega}}_{s;i=0,k=2}\right\} \ . \label{eq:p4:beta}
\end{align}
Here, $\mathbb{R}\{\}$ is the real part of a complex number. Spectral quantities are indicated by $\widetilde{\cdot}$. Quantities $\widetilde{\bm{u}^2}$ and $\widetilde{\bm{\omega}}_s$ are the Fourier- and Chebyshev-transformed kinetic energy $\bm{u}^2=\bm{u}^2(x,y,z,t)$ and the streamwise vorticity $\bm{\omega}_s=(\nabla\times\bm{u}(x,y=0,z,t))\cdot \hat{\bm{e}}_s$ at midplane, respectively. Fourier modes in $x$ and $z$ are indexed by $i$ and $k$. Chebyshev modes in $y$ are indexed by $j$. Thus, $\alpha(t)$ and $\beta(t)$ are the time-dependent real parts of the second Fourier mode along $z$ of the mean kinetic energy and midplane streamwise vorticity, respectively. This Fourier mode corresponds to the dominant oscillating mode of the standing wave with wavelength $\lambda=20$, shown by the $z$-profile in Fig.~\ref{fig:p4:stripeOrbit}b. Along the orbit of $OWR_1$ with period $T=225.4$, $\alpha(t)$ and $\beta(t)$ oscillate with a phase lag of approximately $T/4$ (Fig.~\ref{fig:p4:edgeOrbit}c). In this projection, the phase portrait of $OSW_{1}$ shows a disentangled elliptical loop along which the dynamics revolves in a clockwise sense (Fig.~\ref{fig:p4:edgeOrbit}d). Consequently, the quantities $\alpha(t)$ and $\beta(t)$ illustrate the oscillatory behaviour of the UPO better than $I(t)$ and $D(t)$. 

\begin{figure}[tb]
        \begin{tikzpicture}
    	\draw (0, 0) node[inner sep=0]{\includegraphics[width=0.99\linewidth,trim={0.9cm 2.0cm 0.5cm 0.0cm},clip]{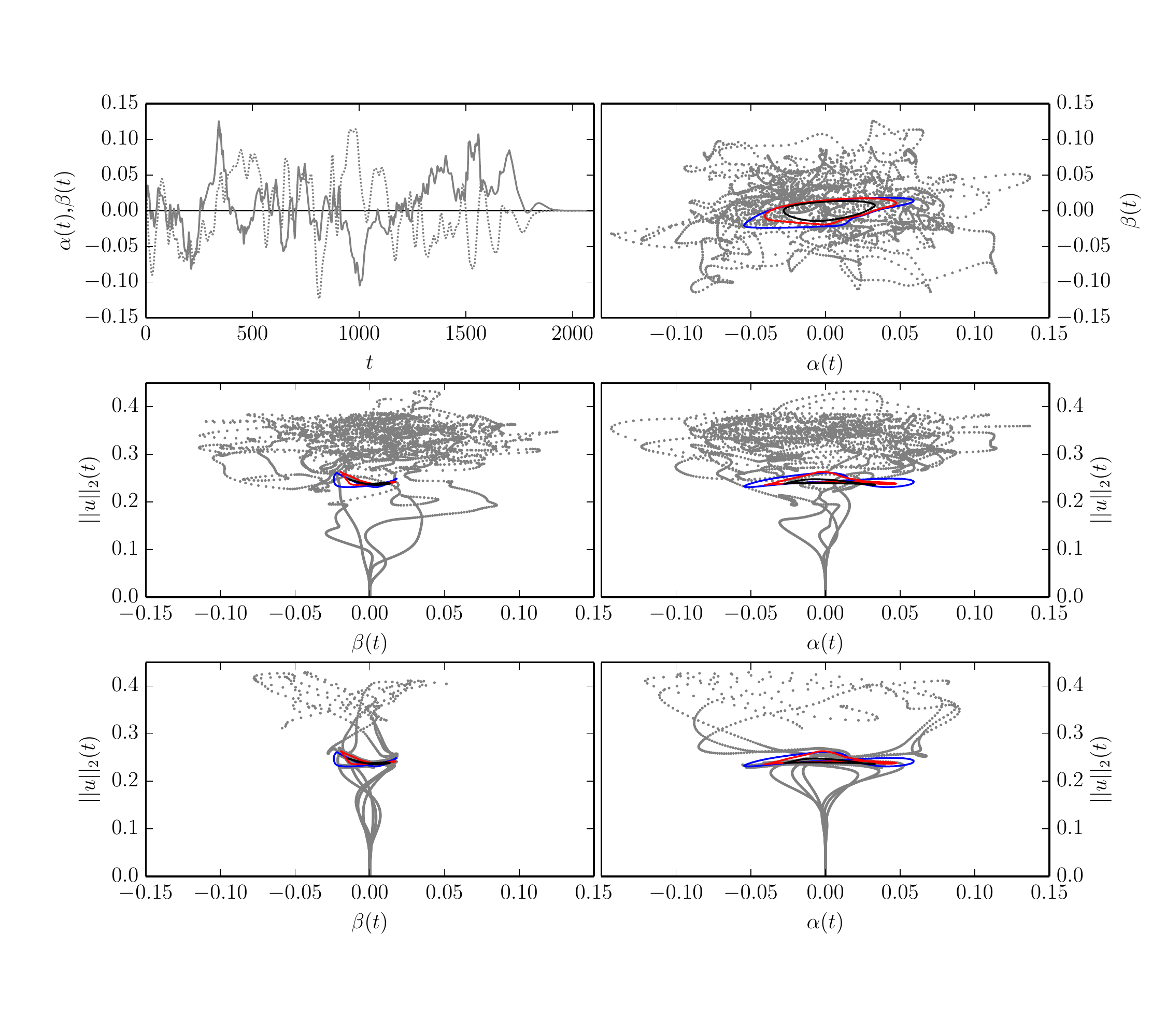}};
    	\draw (-6.9,1.6) node {\textbf{(a)}};
    	\draw (7.0,1.6) node {\textbf{(b)}};
    	\draw (-6.9,-2.2) node {\textbf{(c)}};
    	\draw (7.0,-2.2) node {\textbf{(d)}};
    	\draw (-6.9,-6.4) node {\textbf{(e)}};
    	\draw (7.0,-6.4) node {\textbf{(f)}};
		\end{tikzpicture}
\caption{\label{fig:p4:turbPortrait} Turbulent trajectory in subspace $\mathbb{D}$ (compare with Fig.~\ref{fig:p4:4timeseries}) is plotted in terms of $\alpha(t)$, $\beta(t)$ and $||\bm{u}||_2(t)$. Panel \textbf{(a)} indicates the longest of four simulated time series of $\alpha(t)$ (dotted) and $\beta(t)$ (solid) that exceed $t=1000$ before decaying to laminar flow. All four time series are shown in the phase portraits \textbf{(b-d)} that illustrate that the three periodic orbits mark the lower bound of the turbulent saddle in $||\bm{u}||_2(t)$. Panels \textbf{(e-f)} illustrate trajectories starting from two points along each of three periodic orbits $OSW_{1,2,3}$ that are scaled up and down by $1\%$. Six trajectories from the up-scaled state vectors enter the turbulent saddle. Six trajectories from the down-scaled state vectors approach the laminar solution at $(0,0)$. This demonstrates that all periodic orbits, besides not being edge states, are embedded in the edge of chaos.  }
\end{figure}
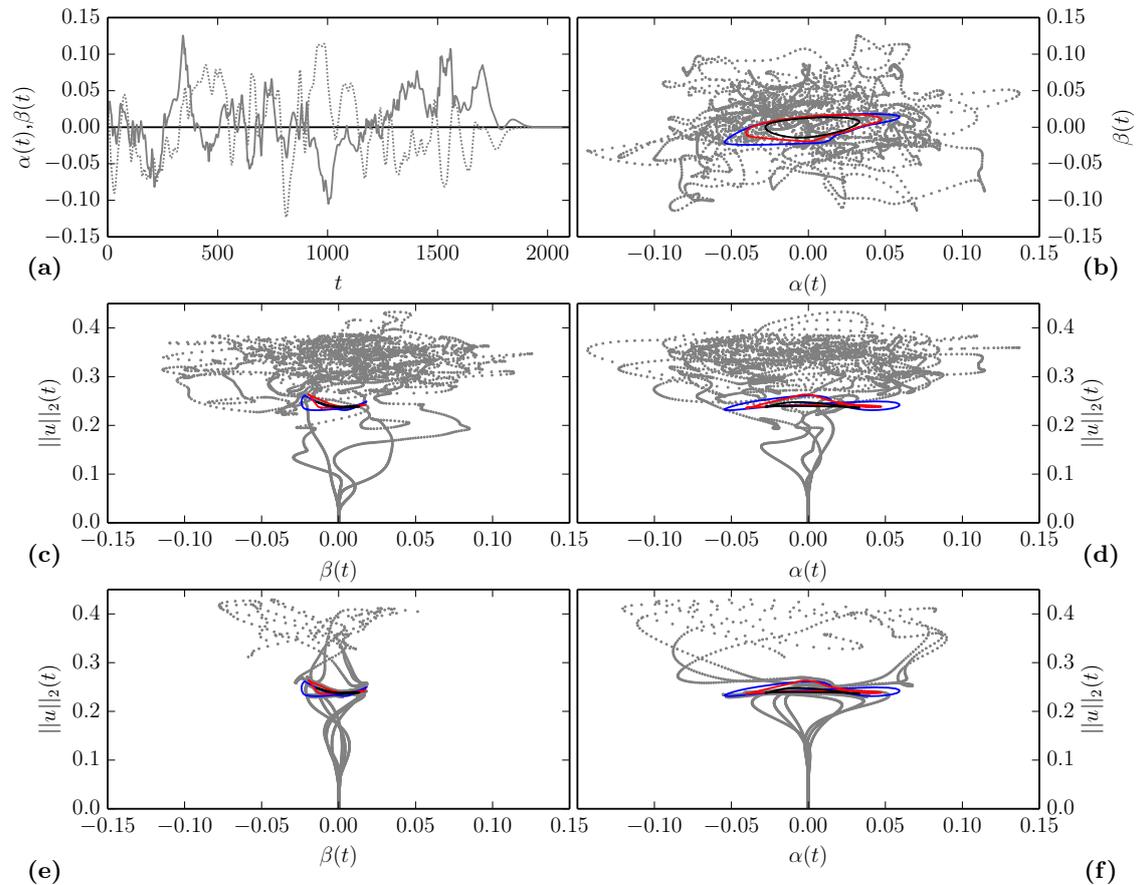

Having identified a good state space projection to display the intrinsic temporal dynamics of the three standing waves with oblique amplitude modulation, we investigate the relevance of these UPOs for the turbulent dynamics and specifically the decay in $\mathbb{D}$. Fig.~\ref{fig:p4:turbPortrait}a shows the turbulent time series, also plotted in Fig.~\ref{fig:p4:4timeseries}d, in terms of $\alpha(t)$ and $\beta(t)$. Here, the phase lag between $\alpha(t)$ and $\beta(t)$ is not constant, varying between approximately in phase and out-of phase. We have computed four transiently turbulent trajectories in subspace $\mathbb{D}$ that all exceed $t=1000$ before decaying to laminar flow. Projecting them onto $\alpha(t)$, $\beta(t)$ and $||\bm{u}||_2(t)$, we find an unstructured cloud of points (Fig.~\ref{fig:p4:turbPortrait}c-d). Yet, the phase portrait clearly indicates the role of the three UPOs for the turbulent dynamics: In the $\alpha$-$\beta$ plane, the orbits are located in the center of the cloud implying that the orbits capture the turbulent mean of $\alpha(t)$ and $\beta(t)$. 

A three-dimensional projection with $\alpha(t)$, $\beta(t)$ and $||\bm{u}||_2(t)$ as coordinates shows that the orbits mark the lower bound in kinetic energy of the turbulent saddle in state space. Prior to decay, the state trajectories pass close by $OSW_{1,2,3}$ in these projections (Fig.~\ref{fig:p4:turbPortrait}). That decaying trajectories transiently visit the periodic orbits is plausible because the periodic orbits are embedded in the edge of chaos. When multiplying state vectors along the orbits $OSW_{1,2,3}$ with factors of either $1.01$ or $0.99$, the associated trajectories either intrude the turbulent saddle or decay to laminar flow, respectively (Fig.~\ref{fig:p4:turbPortrait}e,f). This provides evidence that all periodic orbits are indeed embedded in the edge of chaos, besides not being edge states.

\section{Discussion}
\label{sec:p4:discussion}

We study the spatio-temporal dynamics of oblique stripe patterns in PCF at $\mathrm{Re}=350$ within a numerical dynamical systems analysis. Patterns in four different symmetry subspaces are simulated using DNS in doubly periodic domains of different size and with additionally imposed discrete symmetries. Imposing symmetries reduces the degrees of freedom in the flow and thereby the spatio-temporal complexity of the turbulent patterns. It is shown that inversion symmetric oblique stripe patterns at orientation $\theta=24^{\circ}$ with pattern wavelengths of $\lambda=40$ and $\lambda=20$ are transiently sustained in a minimal tilted domain. Using edge-tracking in the symmetry subspaces allowing patterns of $\lambda=40$, no steady equilibrium or periodic orbit is approached and edge-tracking suggests a chaotic attractor in the edge of chaos. In the most confined symmetry subspace, edge-tracking approaches a slow periodic orbit with a pattern wavelength $\lambda=20$ and a period $T=225.4$. This unstable periodic orbit represents a standing wave oscillation that modulates wavy velocity streaks obliquely in space. The oscillation exchanges regions of high-amplitude wavy streaks with regions of low-amplitude near-straight streaks in time. Numerical continuation of the periodic orbit towards lower $\mathrm{Re}$ indicates a solution branch that undergoes two folds such that three periodic orbits coexist at $\mathrm{Re}=350$. All three periodic orbits represent standing waves, are weakly unstable and embedded in the edge of chaos but have more than one unstable eigenvalue, unlike edge states. Interestingly, the edge-tracking trajectory approaches the state space neighborhood of the most unstable of the three discussed periodic orbits. This suggests that $OSW_1$ is embedded in a more complicated attractor in the edge of chaos, potentially a torus.

While oblique stripe patterns are typically not observed at a pattern wavelength of $\lambda=20$ in experiments or numerical simulations in large extended domains, the three periodic orbits with $\lambda=20$ discussed in this article show features that are dynamically relevant for oblique stripes at larger pattern wavelengths. We highlight three aspects:

First, the standing wave modulation involves skewing and bending effects of wavy velocity streaks (Fig.~\ref{fig:p4:stripeOrbit}d). Such skewing and bending effects are characteristic features that were also found in the analysis of the recently described stripe equilibrium solution \cite{Reetz2019a} and in the detailed analysis of snakes-and-ladders bifurcation structures of spanwise localized wavy velocity streaks \cite{Gibson2016}. The maximum degree of skewing or bending of wavy velocity streaks against the streamwise direction is related to the maximum degree to which periodic domains resolving wavy velocity streaks can be tilted. The angle of tilt of the periodic domain reflects the angle of obliqueness of the stripe pattern. Thus, the range of skewing and bending of wavy streaks that is observed along the UPOs discussed here may suggest the range of orientation angles at which stripes can form. 

Second, the discussed UPOs demonstrate the existence of large-scale waves that obliquely modulate the amplitude of wavy velocity streaks. We expect such waves to also exist at a pattern wavelength of $\lambda=40$. The standing waves show slow dynamics with periods on the order of the viscous diffusion time scale, $T\sim T_d=h^2/\nu$, which due to the nondimensionalization is also $T_d=\mathrm{Re}$. Oblique standing waves with phase velocity (\ref{eq:p4:phasevelocity}) that approximately resonate with the viscous diffusion time scale, $T\approx T_d$, thus imply the approximate dispersion relation
\begin{equation}\label{eq:p4:resonance}
c\approx \frac{\lambda}{\mathrm{Re} \sin(\theta)} \ .
\end{equation}
Assuming $c=\pi^{-1}=0.32$, compatible with the observed values (Fig.~\ref{fig:p4:bifDiag}d), yields the same approximate relation that has previously been suggested based on a mean flow analysis of oblique stripe patterns at a pattern wavelength $\lambda=40$ \cite{Barkley2007}. The cited relation stems from the observation that ``a non-trivial [mean] flow is maintained in the laminar regions by a balance between viscous diffusion and nonlinear advection'' \cite{Barkley2007}. In analogy, dispersion relation (\ref{eq:p4:resonance}) describes an approximate resonance between viscous diffusion and nonlinear wave propagation. We find $c\approx 0.22$ for $OSW_1$ with $\lambda=20$ and $\theta=24^{\circ}$. Continuation of $OSW_1$ in $\lambda$ does not reach $\lambda=40$. However, the observed function $c(\lambda,\theta)$ extrapolates to $\lambda\approx 40$ with a phase velocity of $c\approx \pi^{-1}$ (Fig.~\ref{fig:p4:bifDiag}d).

Third, the UPOs have the same symmetries as the equilibrium solution with pattern wavelength $\lambda=20$ that gives rise to the stripe equilibrium with wavelength $\lambda=40$ via a symmetry-breaking bifurcation, as discussed in \cite{Reetz2019a}. A similar symmetry-breaking bifurcation may exist along the UPO branches discussed here. Such a bifurcation would create a UPO with a pattern wavelength of $\lambda=40$, the typically observed wavelength in experiments.
\\
\\
\noindent \textbf{Acknowledgments}\\
This work was supported by the Swiss National Science Foundation (SNF) under grant no. 200021-160088.\\


%
\end{document}